# Characterizing the Host Galaxies and Delay Times of Ca-rich Gap Transients versus 91bg-like SNe and Normal Type Ia SNe

Peter Scherbak[1], Abigail Polin[2], Mansi Kasliwal[1], Kishalay De[3,4], Peter Behroozi[5], Dave Cook[6], and W. V. Jacobson-Galán[1,7]

[1] California Institute of Technology, Astronomy Department, Pasadena, CA 91125, USA; pscherba@caltech.edu
[2] Purdue University, Physics & Astronomy, West Lafayette, IN 47905, USA
[3] Department of Astronomy and Columbia Astrophysics Laboratory, Columbia University, 550 W 120th St. MC 5246, New York, NY 10027, USA
[4] Center for Computational Astrophysics, Flatiron Institute, 162 5th Ave., New York, NY 10010, USA
[5] University of Arizona, Department of Astronomy and Steward Observatory, Tucson, AZ 85721, USA
[6] Caltech/IPAC, 1200 E. California Boulevard, Pasadena, CA 91125, USA

## Abstract

Calcium-rich gap transients are a faint, fast-evolving class of supernovae that show strong nebular Ca emission lines. Their progenitor systems are uncertain, but they are often associated with old and quiescent host galaxies. In this work, we compare the properties of the hosts of hydrogen-poor Ca-rich gap transients to the hosts of 3 other classes of supernova (SNe): normal Type Ia, 91bg-like, and Type II. We use data from the Zwicky Transient Facility Census of the Local Universe (CLU) experiment to build up our 4 SNe samples and identify the host galaxies. A combination of precomputed host properties from the CLU catalog and those derived from spectral energy distribution fitting are used to characterize each host's stellar mass, star formation rate, and specific star formation rate (sSFR). We find that the hosts of Ca-rich gap transients and 91bg-like SNe occupy a similar parameter space of mass and sSFR, and are more massive and quiescent compared to the hosts of Type Ia and Type II SNe. Additionally, we construct delay time distributions (DTDs) for our 4 samples, finding that Ca-rich gap transients and 91bg-like SNe have the longest peak delay times $\sim 10^4$ Myr, compared to the peak delay times of Type Ia SNe ($\sim 10^3$ Myr) and Type II SNe ($\sim 10$ Myr). The similarity of host environment and DTDs for Ca-rich gap transients and 91bg-like SNe motivates further analysis of the relationship of these two transient classes.

## 1. Introduction

Calcium-rich gap transients are a type of hydrogen-poor supernova (SN) with unique characteristics and unknown progenitors. They are part of a broader class of Ca-rich transients that include hydrogen and that may result from core-collapse supernova (Milisavljevic et al. 2017; Jacobson-Galán et al. 2020a; Das et al. 2023), which are not the focus of this work. Calcium-rich gap transients are fainter and faster-evolving than Type Ia SNe (SNe Ia) and core-collapse SNe, and are identified by strong [Ca II] emission in their nebular phase (Perets et al. 2010; Kasliwal et al. 2012; Milisavljevic et al. 2017; De et al. 2020; Jacobson-Galán et al. 2020a). They may produce $\sim 0.1 M_\odot$ of Ca and thereby may help chemically enrich the universe (Perets et al. 2010). The putative hosts of Ca-rich gap transients are often older and quiescent (Dong et al. 2022) and, curiously, the site of the explosion is commonly located at large offsets from the host (Perets et al. 2010; Kasliwal et al. 2012; Foley 2015; Lunnan et al. 2017; De et al. 2020). This offset is a real effect, not an observational bias, with the rates of these events between one third to equal that of the volumetric rate of SNe Ia (Frohmaier et al. 2018). Additionally, stellar populations have not been detected at the location of most Ca-rich gap transients, suggesting their progenitors may have traveled a great distance from their parent stellar population (Lyman et al. 2014; Foley 2015; Lunnan et al. 2017; De et al. 2020) although this is uncertain (Perets & Beniamini 2021).

Several progenitor models have been put forth to explain Ca-rich gap transients, including helium (He) detonation on a WD (Bildsten et al. 2007; Shen & Bildsten 2009; Waldman et al. 2011; Dessart & Hillier 2015), double detonation of a He shell on a WD (Polin et al. 2019, 2021; Touchard-Paxton et al. 2025), a WD disruption/merger (Zenati et al. 2019, 2023), and the core-collapse of an ultra-stripped massive star (Tauris et al. 2015), but none have been fully validated (Dong et al. 2022). Some Ca-rich gap transients show double-peaked light curves, which may arise from a progenitor star with an extended envelope (Jacobson-Galán et al. 2022; Ertini et al. 2023) or from $^{56}$Ni in the outer layers of the ejecta (De et al. 2018).

[7] NASA Hubble Fellow.

Similarities between Ca-rich gap transients and 91bg-like SNe may exist. 91bg-like SNe are a subclass of SNe Ia that are fainter and redder than normal SNe Ia, and also show faster decline times than normal SNe Ia (Filippenko et al. 1992; Leibundgut et al. 1993; Turatto et al. 1996). 91bg-like SNe are preferentially found in stellar populations with little recent star formation (characteristic of elliptical or lenticular morphologies (Li et al. 2011; Senzel et al. 2025)), suggesting they explode at much later times (i.e., have longer delay times) than normal SNe Ia (Chakraborty et al. 2024), which are found in stellar populations from the very young to very old (Panther et al. 2019). On average, the hosts of normal SNe Ia tend to be bluer/younger than those of 91bg-like SNe (Barkhudaryan et al. 2019; Hakobyan et al. 2020; Perrefort et al. 2020; Burgaz et al. 2025). Additionally, Barkhudaryan et al. (2019) found that 91bg-like SNe show no evidence of prompt progenitors. The subluminous and fast-evolving nature of 91bg-like SNe, their preference for older/quiescent hosts, larger offsets compared to SNe Ia (De et al. 2020), and spectroscopic similarities to some Ca-rich gap transients (De et al. 2020; Jacobson-Galán et al. 2020b) inspire further comparison of these two transient classes.

The usefulness of comparing delay times of normal SNe Ia and 91bg-like SNe motivates construction of delay time distributions (DTDs) of Ca-rich gap transients in comparison to other event types. A DTD shows the transient rate versus time that follows a burst of star formation (Maoz et al. 2012; Heringer et al. 2019; Freundlich & Maoz 2021). Based on the evolutionary timescales preceding the transient, different progenitor scenarios for a transient predict different DTDs (Greggio 2005; Maoz et al. 2012; Heringer et al. 2019). Transients with short (Myr) delay times (e.g., Type II supernova) are linked to massive stars (e.g., Zapartas et al. 2017), whereas those with time delays of Gyr (e.g., SNe Ia) are linked to less massive stars or less common mechanisms (P. Behroozi 2025, in preparation). DTDs can be estimated from the fraction of transients that occur in star-forming versus quiescent galaxies (e.g., Zheng & Ramirez-Ruiz 2007) because, using ensemble averages, it is possible to constrain star formation histories (SFHs) better than could be done for an individual galaxy (e.g., Pacifici et al. 2016; Behroozi et al. 2019).

We leverage transient discoveries with the Zwicky Transient Facility (ZTF; Bellm et al. 2019; Graham et al. 2019) to compare the host galaxies of different transients, especially through the construction of DTDs using host properties plus detections. Section 2 describes the survey and the 4 samples we model: Ca-rich gap transients, 91bg-like SNe, normal SNe Ia and Type II SNe. Section 3 details the characterization of the host galaxies through both spectral energy distribution (SED) modeling with PROSPECTOR (Leja et al. 2017; Johnson et al. 2021), and also with data from the Census of the Local Universe (CLU) catalog. In Section 4, we compare the properties of the host galaxies of the different transient classes, and present DTDs constructed from the masses and star-forming statuses of the hosts. We discuss our results and conclude in Section 5.

## 2. Observations: The ZTF-CLU Experiment

The ZTF is a optical time-domain survey that uses the Palomar 48 inch Schmidt telescope and surveys the Northern Hemisphere sky in three bands (Bellm et al. 2019; Graham et al. 2019). The ZTF-CLU experiment (De et al. 2020) is designed to classify transients coincident with galaxies in the CLU catalog (Cook et al. 2019). The spatial cross-match radius between transient and galaxy is set to $3 \times D_{25}$ (the isophotal major axis containing 25 of the galaxy's total light) or 280$''$ when $D_{25}$ is not available (corresponding to 270 kpc at a distance of 200 Mpc) (De et al. 2020). The CLU catalog extends to 200 Mpc and consists of previously identified galaxies as well as galaxies found with an emission-line (H$\alpha$) survey by the Palomar 48 inch telescope (Cook et al. 2019). As of 2023, the CLU catalog contains ∼270,000 galaxies in total. The CLU catalog includes estimates of a galaxy's stellar mass and star formation rate (SFR), with the latter computed via FUV fluxes.

This work mainly uses a sample of eight hydrogen-poor Ca-rich gap transients identified in the ZTF-CLU experiment from 2018 June 1 to 2019 September 30 (De et al. 2020) (our Gold sample). De et al. (2020) finds that the average Ca-rich gap transient is detectable out to 150 Mpc for a flux limit in $r$ band of 20.0 mag (the target limiting magnitude of the ZTF-CLU experiment). We also identify nineteen Ca-rich transients detected earlier (mostly before 2018, by other surveys) and modeled in Dong et al. (2022), which we designate as our Silver Sample (Section 3.2). However, only a fraction of these nineteen can be conclusively identified as Ca-rich *gap* transients due to a lack of photometry near peak light which is required for the identification. Additionally, some of the events in the Dong et al. (2022) sample show hydrogen, which represents a different class of Ca-rich transient. We therefore cross match the selection of Dong et al. (2022) with known Ca-rich gap transients compiled in De et al. (2020), which used the same cuts in generating our Gold sample, when comparing the host properties.

We identified other classes of transients through the Fritz data platform.[8] Over the same period as the Ca-rich gap transients were detected, ZTF detected about 290 normal SNe

[8] https://fritz-marshal.org/

Ia, 270 Type II SNe, and 30 91bg-like SNe as part of the ZTF-CLU experiment. Together these comprise 4 samples of transients, for which we analyze the host galaxies.

## 3. Determining Properties of Host Galaxies

We identify the host galaxies for all transients by querying the CLU catalog, and estimate each host's SFR and stellar mass $M_*$ through two main methods. All quoted values are median values. For the sample of Ca-rich gap transients and the sample of 91bg-like SNe, we perform host galaxy modeling in PROSPECTOR, which infers stellar population properties based on photometric data (Leja et al. 2017; Johnson et al. 2021). For all 4 samples we obtain $M_*$ and SFR directly from the CLU catalog, where the quoted SFRs were calculated primarily via the galaxy's FUV flux (from the GALEX mission (Martin et al. 2005; Bianchi et al. 2014)) and additionally are dust-corrected using the W4 flux (Wright et al. 2010), making the derived values more accurate.

In all cases, we calculate the host's specific star formation rate (sSFR) as the ratio of SFR to $M_*$. We adopt the criteria that sSFR $> 10^{-11}\,\mathrm{yr}^{-1}$ implies a star-forming galaxy, whereas sSFR $< 10^{-11}\,\mathrm{yr}^{-1}$ implies a quiescent galaxy (Fontana et al. 2009).

### 3.1. Prospector Analysis

For our most rigorous analysis, we characterize the host galaxies using the Python tool PROSPECTOR, which infers stellar population properties based on photometric data (Leja et al. 2017; Johnson et al. 2021) and uses FLEXIBLE STELLAR POPULATIONS (Conroy et al. 2009; Conroy & Gunn 2010a) and PYTHON-FSPS (Foreman-Mackey et al. 2014).

Where available, we use photometry from the following surveys in our fitting. We use optical data from the Sloan Digital Sky Survey Data Release 17 in *u*, *g*, *r*, *i*, *z* bands (Abdurro'uf et al. 2022). We retrieve data (W1, W2, W3, W4 bands) from the AllWISE catalog in the NASA/IPAC INFRARED SCIENCE ARCHIVE (IRSA; Wright et al. 2010; Mainzer et al. 2011). WISE magnitudes are converted into the AB system using Wright et al. (2010). We use elliptical aperture photometry measurements when available, as these do a better job of capturing extended source brightness (Cutri et al. 2013). We retrieve data (*J*, *H*, *K* bands) from the 2MASS All-Sky Extended Source Catalog (Skrutskie et al. 2006) in IRSA and also convert to the AB system. Finally, we use data (NUV, FUV bands) from GALEX (Martin et al. 2005), accessed through the GALEX GR6/7 Data Release[9] in Mikulski Archive for Space Telescopes (MAST). We use standard SDSS, WISE, 2MASS and GALEX transmission curves from SEDPY (Johnson 2019).

We retrieve redshifts from the NASA/IPAC Extragalactic Database (NED),[10] and correct SDSS magnitudes for galactic dust extinction using NED (Schlafly & Finkbeiner 2011). We adopt 10% uncertainties on all spectral fluxes ("maggies" in PROSPECTOR) as most fluxes have small observational uncertainties, but systematic uncertainties can exist (Conroy et al. 2009; Conroy 2013).

We adopt many of the same priors as Dong et al. (2022), who modeled the host galaxies of nineteen Ca-rich transients detected from 2000 to 2019. We use a Chabrier initial mass function (Chabrier 2003), and a Milky Way Extinction law parameterized by Cardelli et al. (1989). We include a model of dust emission from Draine & Li (2007) which describes the PAH thermal emission features, but we keep the dust model's parameters fixed. We also include a nebular continuum in our model and impose a 2:1 ratio on the amount of dust attenuation between the young and old stellar populations (Calzetti et al. 2000). We do not assume a stellar mass–stellar metallicity relationship.

Instead of a parametric delayed-$\tau$ model for SFH, we use PROSPECTOR's CONTINUITY_SFH_MODEL, which is less biased and produces better error estimates (Leja et al. 2019). We fit to four parameters in this model: metallicity ($Z$), dust extinction ($A_V$), total mass formed, and SFR ratios between adjacent temporal bins. Following Dong et al. (2022), we use eight temporal bins, as the fitting should not use less than four bins (Leja et al. 2019). The first two bins are fixed to 0–30 and 30–100 Myr, while the last bin is calculated as the age of the universe at the galaxy's redshift, using the cosmology calculator of Wright (2006). We ran PROSPECTOR using the DYNESTY sampler (Speagle 2020).

We convert the total mass formed to stellar mass ($M_*$) using a fair sample of the posterior. We transform the total mass and SFR ratios into SFR values, and then calculate the host's sSFR by dividing the SFR in the most recent age bin by the surviving stellar mass.

### 3.2. Host Properties: PROSPECTOR versus CLU Catalog

For the eight Ca-rich gap transients in our Gold sample, the results of the PROSPECTOR fitting for $M_*$, SFR, and sSFR are shown in Table 1. Based on the criterion discussed above, we find that all eight hosts are quiescent. Three hosts lack a detection in one of the four surveys for which we use photometry—however, these three all had at least nine photometry points, which Dong et al. (2022) found to be the minimum required for a confident analysis using the CONTINUITY_SFH_MODEL model.

The CLU catalog contains $M_*$ estimates for all eight hosts, and SFR estimates (from FUV fluxes) for five of the eight. We

[9] http://galex.stsci.edu/gr6/

[10] The NASA/IPAC Extragalactic Database (NED) is funded by the National Aeronautics and Space Administration and operated by the California Institute of Technology.

**Table 1**
Properties of the Host Galaxies of Eight Ca-rich Gap Transients (Our Gold Sample) Modeled in PROSPECTOR

| Transient Name | Host Name | log $(M_*/M_\odot)$ | SFR ($M_\odot$ yr$^{-1}$) | log(sSFR) (yr$^{-1}$) | Host Morphology |
|---|---|---|---|---|---|
| ZTF 18aayhylv/SN 2018ckd | NGC 5463 | $11.25^{0.03}_{0.09}$ | 0.0038 | $-13.78^{0.46}_{0.81}$ (q) | S0 |
| ZTF 18abmxelh/SN 2018lqo | CGCG 224-043 | $10.86^{0.02}_{0.03}$ | 0.012 | $-12.69^{0.07}_{0.08}$ (q) | E |
| ZTF 18abttsrb/SN 2018lqu | WISEA J155413.91+133102.4 | $9.93^{0.02}_{0.02}$ | 1.3e-5 | $-15.25^{1.61}_{4.37}$ (q) | E |
| ZTF 18acbwazl/SN 2018gwo | NGC 4128 | $10.67^{0.04}_{0.06}$ | 0.0050 | $-13.00^{0.14}_{0.34}$ (q) | S0 |
| ZTF 18acsodbf/SN 2018kjy | NGC 2256 | $11.40^{0.02}_{0.02}$ | 0.0057 | $-13.61^{0.16}_{0.25}$ (q) | E |
| ZTF 19aaznwze/SN 2019hty | WISEA J125534.50+321221.5 | $10.40^{0.01}_{0.02}$ | 0.009 | $-12.37^{0.09}_{0.11}$ (q) | E |
| ZTF 19abrdxbh/SN 2019ofm | IC 4514 | $10.70^{0.08}_{0.09}$ | 0.097 | $-11.71^{0.47}_{1.02}$ (q) | spiral |
| ZTF 19abwtqsk/SN 2019pxu | WISEA J051011.32-004702.5 | $10.55^{0.09}_{0.08}$ | 0.015 | $-12.60^{0.87}_{1.93}$ (q) | spiral |

**Note.** Values are generated from the posterior distributions using the CONTINUITY_SFH_MODEL (fitting for total mass, SFR ratios, $A_V$, and $Z$). Total mass is transformed into surviving stellar mass $M_*$. SFR ratios are used to generate the SFR bins. The sSFR is calculated using the most recent SFR bin and $M_*$, and (q) refers to quiescent (the median sSFR below $<10^{-11}$ yr$^{-1}$). Morphologies are from De et al. (2020).

find a significant discrepancy between the $M_*$ from PROSPECTOR modeling and $M_*$ from the CLU catalog, the difference ranging from 0.2 to 0.8 dex. The difference may be because early-type galaxies have a different mass to light ($M/L$) ratio (Leroy et al. 2019), as galaxies with low sSFRs as modeled in PROSPECTOR tend to show higher discrepancies. We also find discrepancies for the SFR, with the CLU catalog suggesting higher values, especially for galaxies that have a low sSFR as modeled in PROSPECTOR. This is likely due to the fact that some elliptical galaxies have higher than expected dust content, which is translated into an artificially high estimate for the SFR (Davis et al. 2014; Simonian & Martini 2017). However, when we compute the sSFR with CLU catalog data, we find that the five hosts with SFR data were still classified as quiescent.

We perform similar analysis for the sample of 91bg-like SNe hosts. With the PROSPECTOR analysis, 90% of the 91bg-like SNe hosts are quiescent whereas using data from the CLU catalog, 60% of the 91bg-like SNe hosts are quiescent. As with the Ca-rich gap transient hosts, PROSPECTOR appears to consistently underestimate the SFR compared to the CLU catalog's value.

Using the CLU catalog data, 12% of SNe Ia hosts are deemed quiescent, and <2% of the Type II SNe hosts are quiescent. Using detections in the Bright Transient Survey, Irani et al. (2022) found that <1% of Type II SNe hosts are quiescent, which appears consistent with our results. We did not perform PROSPECTOR analysis for the entirety of the SNe Ia and the Type II SNe hosts because of the large sample size involved (>500 sources in total), but when a subset of SNe Ia hosts were modeled in PROSPECTOR, we again find a discrepancy where PROSPECTOR analysis favored larger $M_*$ and lower sSFRs. The mass discrepancy decreases for star-forming hosts with a higher sSFR (Figure 1), where we have included additional 91bg-like SNe detected later than 2020 to better demonstrate the trend. We again interpret the discrepancies as the $M/L$ ratio depending on the star formation activity of the galaxy, where low sSFRs should actually be associated with higher $M/L$ ratios. Therefore, the CLU catalog may be undervaluing masses for many of the quiescent hosts we model compared to more sophisticated analysis in PROSPECTOR.

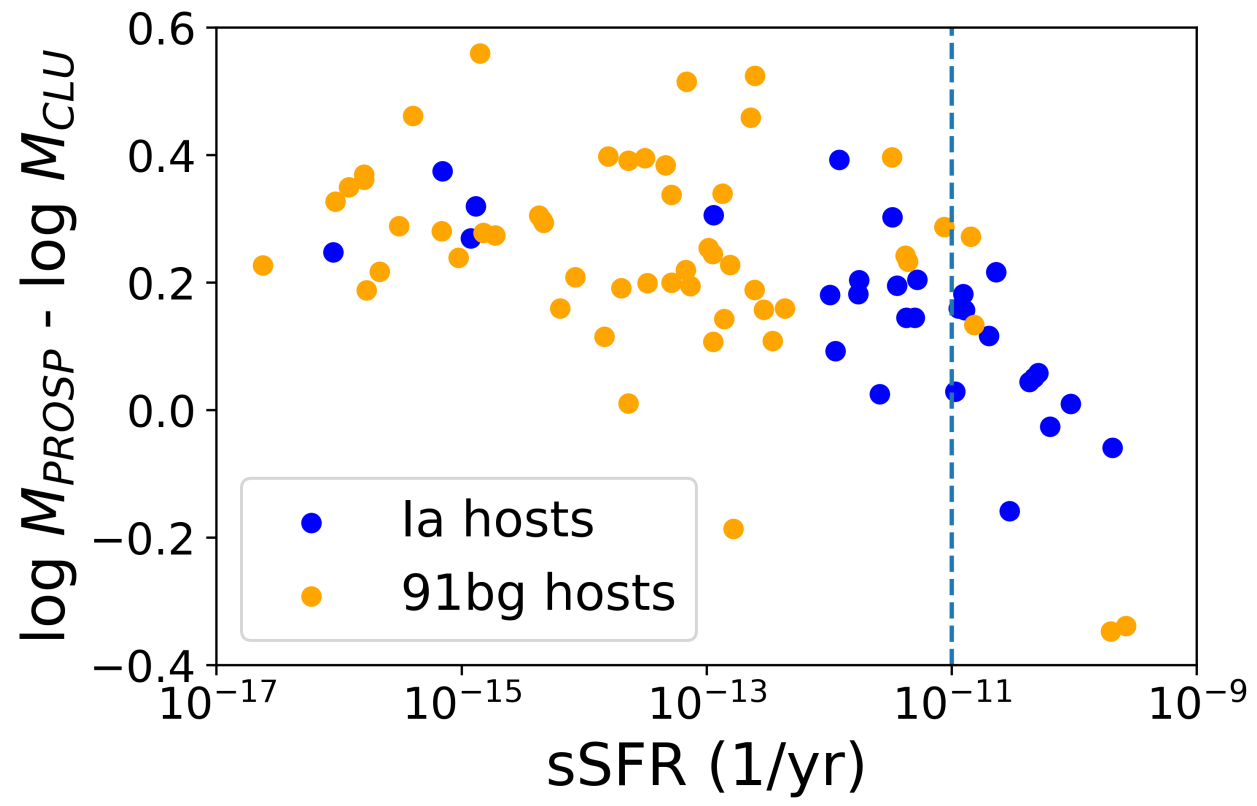

**Figure 1.** The discrepancy between stellar mass as found in PROSPECTOR and the CLU catalog, plotted versus sSFR as found in PROSPECTOR, for 91bg-like SNe and SNe Ia. Additional 91bg-like SNe detected later than 2020 and not in our main sample, but which we modeled in PROSPECTOR, are also included. There appears to be a trend where the discrepancy decreases for star-forming galaxies with sSFR $> 10^{-11}$ yr$^{-1}$ (the boundary marked by the vertical dashed line).

## 4. Results

Figure 2 shows the parameter space of host sSFR and $M_*$ for our transient samples, including Ca-rich gap transients from our Gold sample. Points correspond to the median sSFR and median $M_*$, with errorbars not shown for convenience. The top plot is generated from SFR and $M_*$ estimates using the CLU catalog for all 4 samples, whereas the bottom plot differs in using SFR and $M_*$ from PROSPECTOR modeling for the Ca-rich gap transient and 91bg-like SNe hosts only (see Section 3.2). Although PROSPECTOR analysis is likely more sophisticated than using values from the CLU catalog, the large sample size for SNe Ia and Type II SNe hosts means that we did not perform a direct comparison with PROSPECTOR modeling.

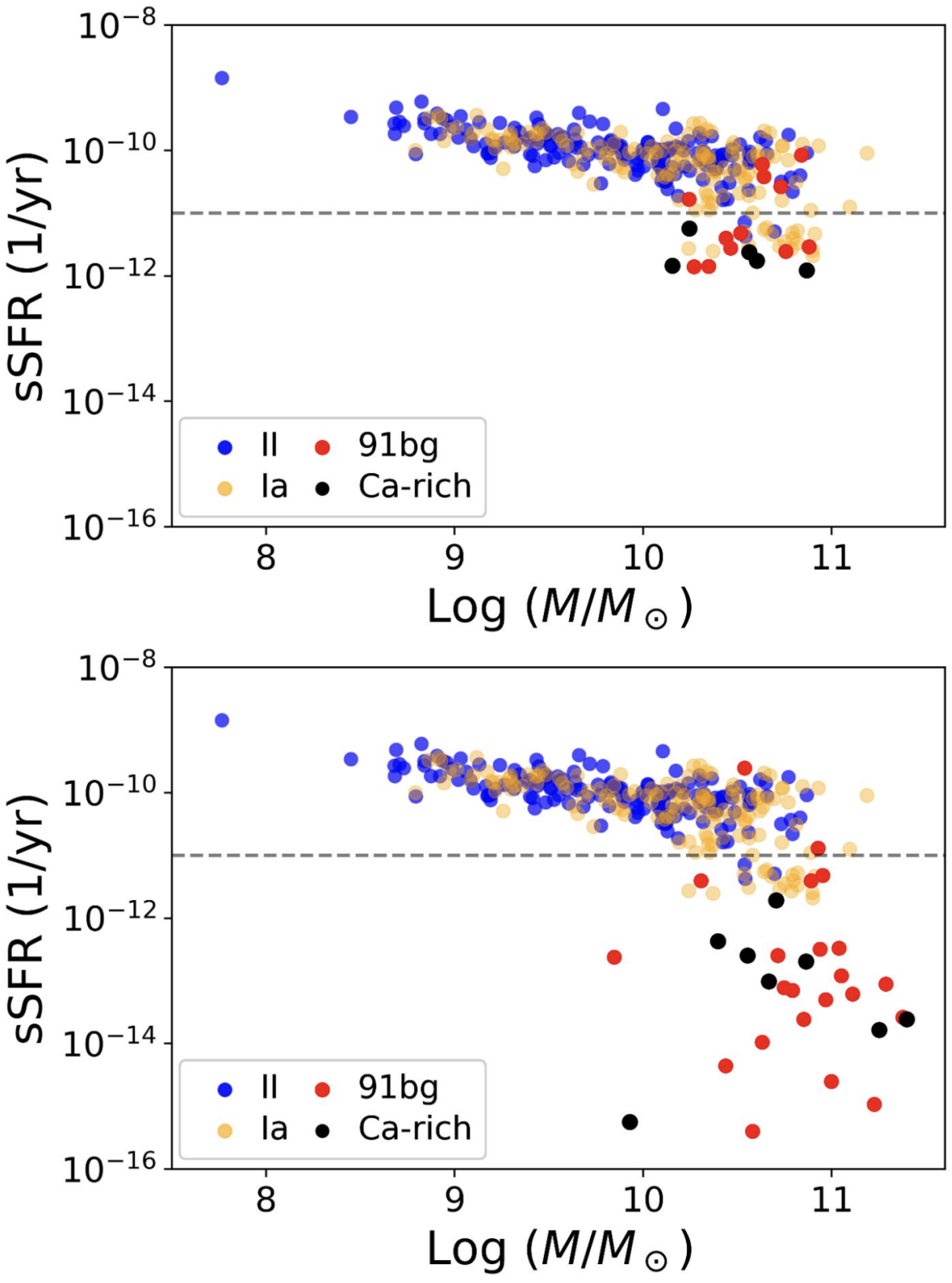

**Figure 2.** The stellar masses and specific star formation rates (sSFR) of the galaxy hosts of 4 classes of transients (Ca-rich gap transients, 91bg-like SNe, normal SNe Ia, Type II SNe). The horizontal line designates the fiducial boundary between star-forming and quiescent galaxies. Top: masses and SFRs are retrieved directly from the CLU catalog for all hosts. Bottom: for the hosts of Type II SNe and SNe Ia, same as top panel. For the hosts of Ca-rich gap transients and 91bg-like SNe, masses and SFRs are estimated using PROSPECTOR modeling.

We find a significant overlap in the parameter space of Ca-rich gap transient and 91bg-like SNe hosts, where they prefer higher mass, quiescent hosts. However, the exact region of overlap differs between the top and bottom plot in that PROSPECTOR modeling prefers more quiescent hosts. In comparison, the hosts of SNe Ia and Type II SNe occupy a range of masses that extends to low mass galaxies, and they tend to be star-forming. Almost none of the Type II SNe hosts are quiescent and a minority of SNe Ia hosts are quiescent, in agreement with Irani et al. (2022). The hosts of 91bg-like SNe being more massive and less star-forming compared to hosts of SNe Ia also agrees well with the literature (Barkhudaryan et al. 2019; Panther et al. 2019; Burgaz et al. 2025; Dimitriadis et al. 2025; Senzel et al. 2025).

Figure 3 includes the addition of our Silver Sample, nineteen Ca-rich transient hosts modeled in Dong et al. (2022) (see Section 2). However, only 10 of these can be conclusively identified as Ca-rich gap transients using the same criteria as De et al. (2020) (low luminosity, hydrogen-poor events that showed strong [Ca II] emission in the nebular phase). Others either have a lack of photometry near peak light to identify them as Ca-rich gap transients, or have photometric properties that differ from Ca-rich gap transients. We therefore differentiate between these two subclasses, labeling the Ca-rich transients that are not gap transients as “other Ca-rich” in Figure 3.

In their PROSPECTOR analysis, Dong et al. (2022) modeled all nineteen hosts with a parametric delayed-$\tau$ model, but only modeled nine with the CONTINUITY_SFH_MODEL model, which requires more photometric points for accuracy. In the latter case, only two of the nine are Ca-rich gap transients. In Figure 3, we therefore plot the results of Dong et al. (2022) when using the parametric delayed-$\tau$ model, although the CONTINUITY_SFH_MODEL may be more accurate. Some of their host properties overlap well with ours, corresponding to quiescent, massive galaxies. However, several Ca-rich gap transients hosts from Dong et al. (2022) are much less massive and are more star-forming than those of our sample. Part of the discrepancy is that parametric models tend to underestimate stellar masses, as Dong et al. (2022) found in their comparison of their models. In addition, Dong et al. (2022) found that six of the nine hosts had higher SFRs when using the parametric models. Together, these two effects might push the hosts of the other Ca-rich gap transients somewhat closer to our sample, which uses the CONTINUITY_SFH_MODEL model for analysis. See also Section 5 for caveats related to the host identification of some of the Dong et al. (2022) sample.

We then use the properties of the hosts to construct DTDs with PYROMETER (P. Behroozi 2025, in preparation), which forward-models expected transient rates by convolving a given DTD with SFHs from the UNIVERSEMACHINE empirical model (Behroozi et al. 2019). The UNIVERSEMACHINE infers SFHs as a function of stellar mass and star-forming status by forward-modeling the SFR as a function of host dark matter halo mass and accretion rate, constrained primarily by observed galaxy number densities and spatial clustering. Relevant for this paper, the SFHs derived in this way are more accurate for early star formation (e.g., more than 2 Gyr ago) than traditional SED-fitting approaches, because they are constrained to match the actual number density of high-redshift galaxies, whereas traditional SED-fitting methods are limited by the color similarity of old stellar populations. Comparisons with advanced SED-fitting approaches (e.g., Pacifici et al. 2016) suggest very similar reconstructions between the UNIVERSEMACHINE and SED-fitting methods over the range of 100 Myr–2 Gyr.

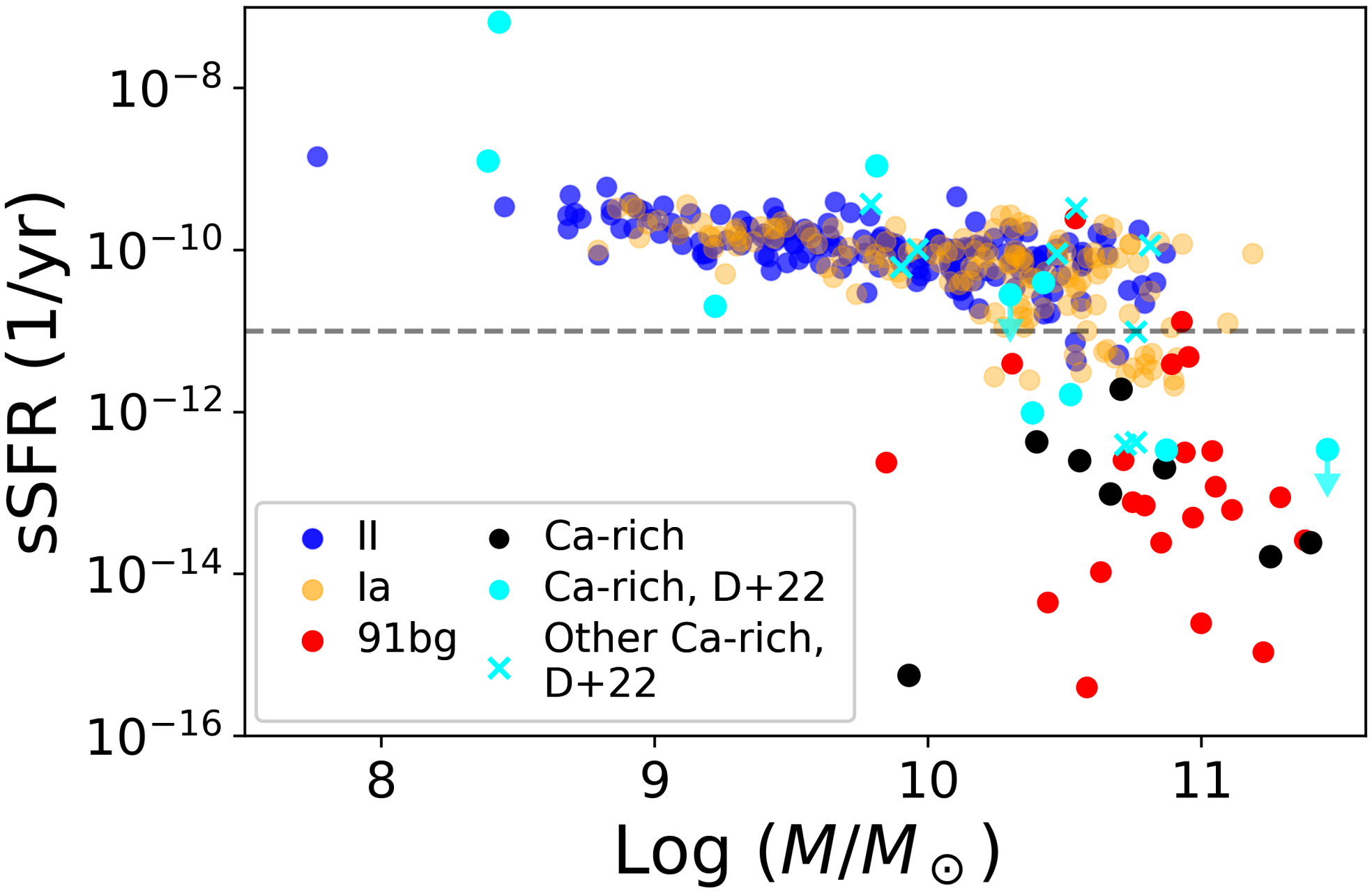

**Figure 3.** Similar to the bottom panel of Figure 2, but with the addition of the Silver Sample, Ca-rich transient hosts identified in Dong et al. (2022) that were modeled in PROSPECTOR. Those from Dong et al. (2022) labeled as "Other Ca-rich" have strong calcium emission, but are not identified as Ca-rich gap transients. The two points with downward arrows indicate upper limits.

To compare with observations, PYROMETER calculates expected event rates as above for a bin in redshift, galaxy stellar mass, and star-forming status, and uses Poisson statistics to compute a likelihood function for the observations given a chosen model. It then uses a Markov Chain Monte Carlo method with a combination of adaptive Metropolis (Haario et al. 2001; Goodman & Weare 2010) steps to compute the posterior for the DTDs that best match the observations. We have found that 150,000 burn-in steps and 150,000 sampling steps were always enough to achieve convergence in the posterior distributions for the observations in this paper.

We assume the DTD is a power law with some index from some minimum delay time $t_{\rm min}$ to a maximum delay time $t_{\rm max}$. To normalize the transient event rate, the sky area and duration of the survey is necessary. We use as inputs the stellar mass and star-forming quiescent/status of hosts, as found in Section 3, which is a boolean input to PYROMETER. The sky area of ZTF is estimated to be ~26470° ($3\pi$ of the sky) for all samples (Masci et al. 2018) and we use the survey duration discussed in Section 2. The area and duration provide a scaling for the DTD event rates, which is approximate, whereas the relative value of event rate between SN types is precise. For the 91bg-like SNe, SNe Ia and Type II SNe samples, we assume a lower redshift bound of 0, and for upper redshift the equivalent to the ZTF-CLU experiment's 200 Mpc extent ($z \sim 0.049$). At 200 Mpc, the CLU limiting magnitude of $r \approx 20$ mag (De et al. 2020) corresponds to detections of absolute magnitude $M \lesssim -16.5$. For Ca-rich gap transients we use the equivalent redshift to 150 Mpc ($z \sim 0.034$), which De et al. (2020) found to be the effective radius of detection. Using instead the upper redshift of the other samples for the Ca-rich gap transient sample produced no significant effect on the final DTDs. Finally, we estimate the minimum host galaxy stellar mass detectable to be about $10^8\,M_\odot$ for both quenched and star-forming galaxies (Cook et al. 2019). Lowering the minimum bound to $10^7\,M_\odot$ produced negligible results on the final DTDs for our samples.

Figures 4 and 5 show posterior distributions of DTDs for Ca-rich gap transients (Gold sample) and 91bg-like SNe generated using host parameters as modeled in PROSPECTOR. Figure 6 shows a comparison for the 4 samples, with DTDs for Ca-rich gap transients and 91bg-like SNe generated both from PROSPECTOR and CLU catalog estimates. In both cases, 91bg-like SNe and Ca-rich gap transients are associated with significantly longer delay times than SNe Ia and Type II SNe, with the median event rate (black line) peaking at longer delay times. Each gray line is an individual DTD demonstrating that power laws of various slopes and various minimum/maximum delay times are consistent with the data. However, the power laws tend to cluster about the median rate. DTDs that have a value of zero at a given time (i.e., a time outside the range $t_{\rm min}$ to $t_{\rm max}$) are not shown. In the case of Type II SNe, the median event rate appears to fall below most of the individual DTDs. This is because the DTDs with zero values at some times bring the median value below the non-zero DTDs.

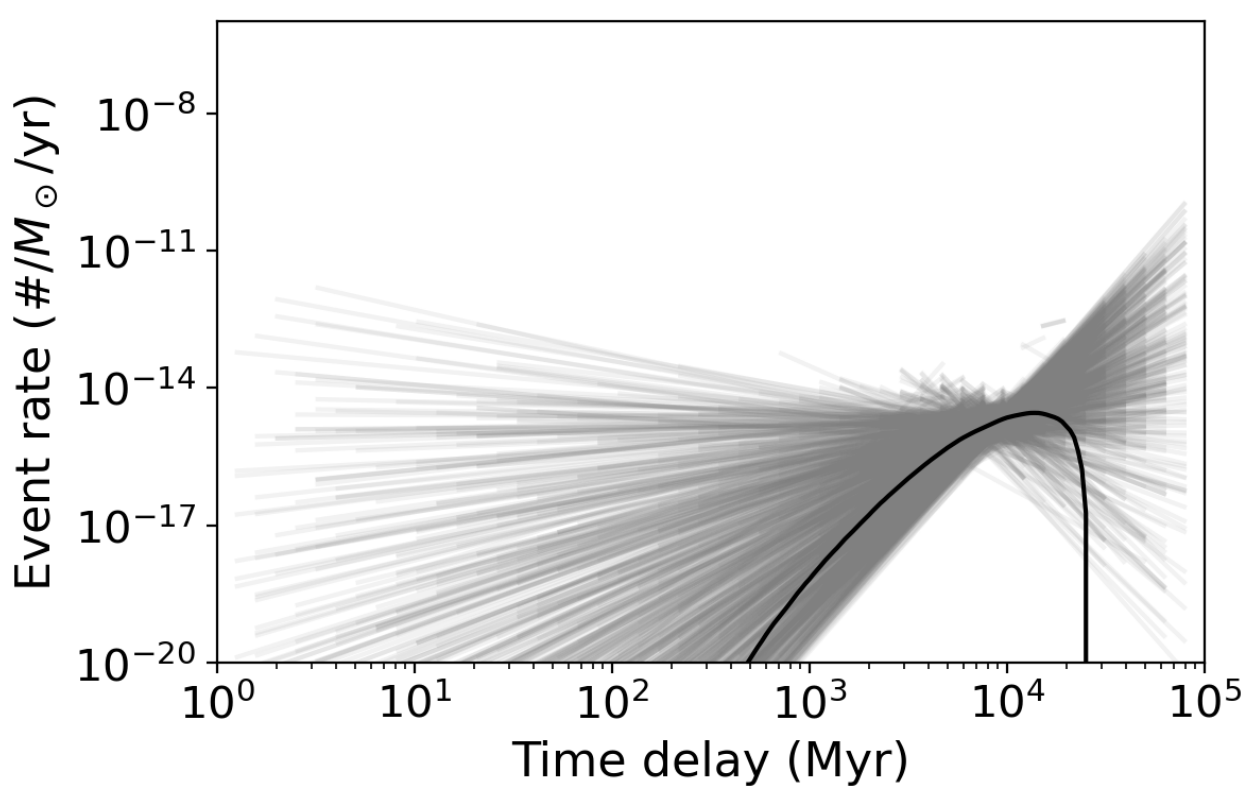

**Figure 4.** Posterior delay time distributions (DTDs) for Ca-rich gap transients, based on the stellar masses and star-forming/quiescent status of their host galaxies. Host properties are determined in PROSPECTOR. The DTDs are assumed to be a power law, and the light gray lines show individual DTDs. The black line is the median event rate from the DTD posterior.

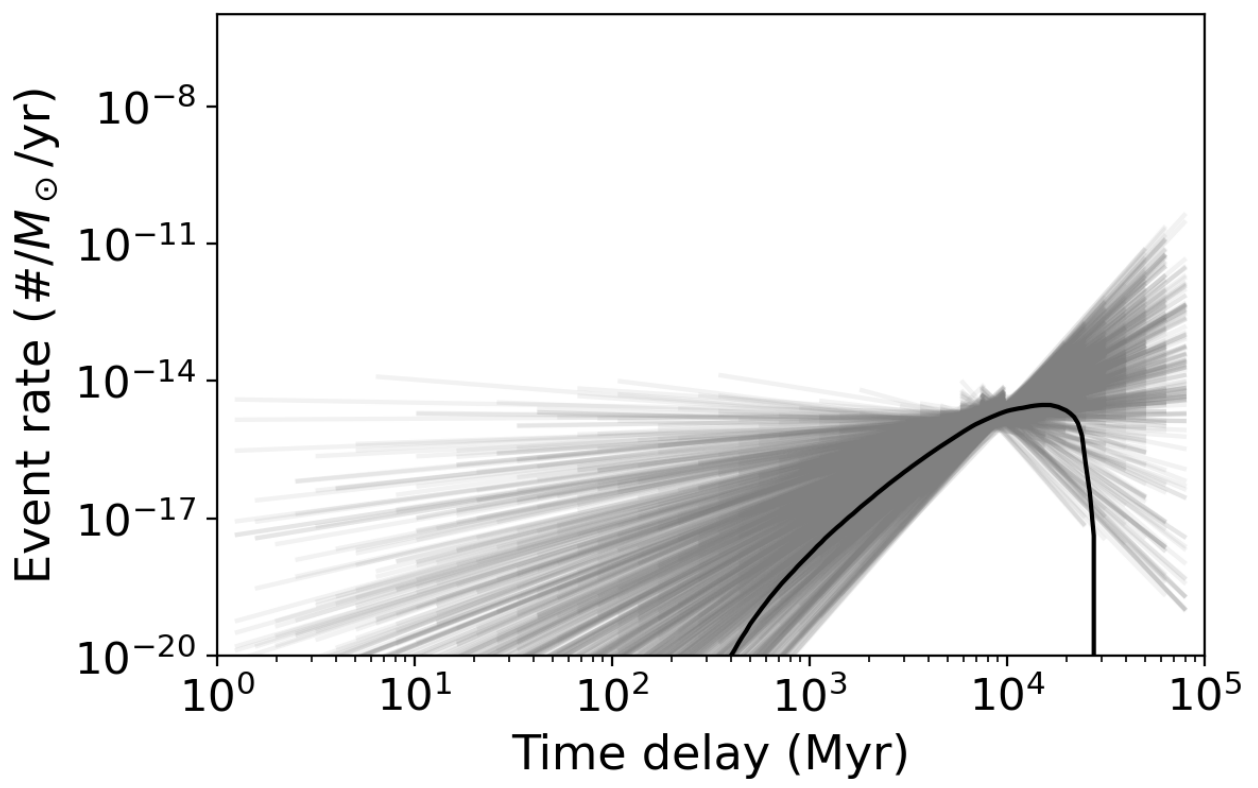

**Figure 5.** Similar to Figure 4, but based on the host galaxies of 91bg-like SNe. Host properties are determined in PROSPECTOR.

## 5. Discussion

This work has characterized the delay times of Ca-rich gap transients using a systematic sample from De et al. (2020). Compared to previous samples of Ca-rich transients, our sample includes only those that are faint/fast and do not include hydrogen. Our results for the delay times of the other three classes of SNe (Section 4) agree with previous work, where 91bg-like SNe, a subset of SNe Ia, have been found to have longer delay times than normal SNe Ia (Barkhudaryan et al. 2019; Panther et al. 2019) and Type II SNe having short delay times (e.g., Zapartas et al. 2017). This paper is the first to calculate DTDs for Ca-rich gap transients, and future work should attempt to connect the DTDs of Ca-rich gap transients to their potential formation histories, as has been done with SNe Ia delay times to distinguish between the single degenerate and double degenerate formation scenario (Maoz et al. 2012).

Comparing the stellar mass-sSFR parameter space of galaxies containing transients is a useful tool (Figures 2 and 3), and we encourage similar analysis for more transient classes. Indeed, such a comparison may be more useful than constructing DTDs because PYROMETER only takes a binary input parameter (star-forming or quiescent) to estimate DTDs, whereas Figures 2 and 3 show the full range of sSFR. Similar to the results of Dong et al. (2022), we find that a large fraction of Ca-rich gap transient hosts are quiescent. However, we find that all hosts in our sample are quiescent, whereas Dong et al. (2022) found that only approximately half were quiescent (although our PROSPECTOR results use a different, potentially more accurate model that requires more host photometry data). We note that our large fraction of quiescent hosts is not surprising given the majority are classified as ellipticals (Table 1). We did not construct DTDs using the Ca-rich transient sample of Dong et al. (2022) because, in addition to about half of their sample not being confirmed as Ca-rich gap transients, the transients they modeled were not detected as part of a systematic survey and the completeness of the sample is unclear.

There are additional caveats, related to the identification of host galaxy, related to some of the hosts of Ca-rich gap transients from Dong et al. (2022) that show the highest sSFR in Figure 3. PTF09dav (Kasliwal et al. 2012) had a fiducial host offset of 40 kpc, and was closer to several faint sources that may be dwarf galaxies (De et al. 2020). PTF11kmb (Lunnan et al. 2017) was located 150 kpc from the most likely host galaxy, and was additionally found in a galaxy group. In contrast, the hosts of our Gold sample had smaller offsets of ≈10−30 kpc, and there was less ambiguity as to the host (De et al. 2020). Therefore, two of the Dong et al. (2022) points (cyan coloring) with highest sSFR in Figure 3 could be wrong if the fiducial host is actually incorrect. iPTF16hgs, which had a peculiar double-peaked light curve, also has one of the most star-forming hosts. However, SN 2018lqo in our Gold sample was also double-peaked (De et al. 2020) and has a quiescent host, so there is no obvious relation between this subclass and host properties.

We have so far treated our Gold sample of Ca-rich gap transients as monolithic. However, De et al. (2020) found that among the eight detections, there are three related spectroscopic subclasses: red Ca-Ib/c objects, with SNe Ib/c-like features and a red color at peak light; green Ca-Ib/c objects, similar to previous but with a green color at peak light; and Ca-Ia objects, with SNe Ia-like features. We compare the host properties of these three subclasses to determine if there is any trend, with the limitation of very small numbers in each of these subclasses. We find, in our PROSPECTOR analysis, that the host of the single Ca-Ia object in our Gold sample has the largest sSFR of all the hosts in the Gold sample, although it is still deemed quiescent. In addition, De et al. (2020) identified two Ca-Ia objects in the literature, whose hosts were modeled in Dong et al. (2022) and were determined to be star-forming galaxies (one of which had the highest sSFR among all the hosts in Dong et al. 2022). We therefore tentatively suggest that Ca-Ia objects are associated with galaxies that are more star-forming, compared to the other two subclasses.

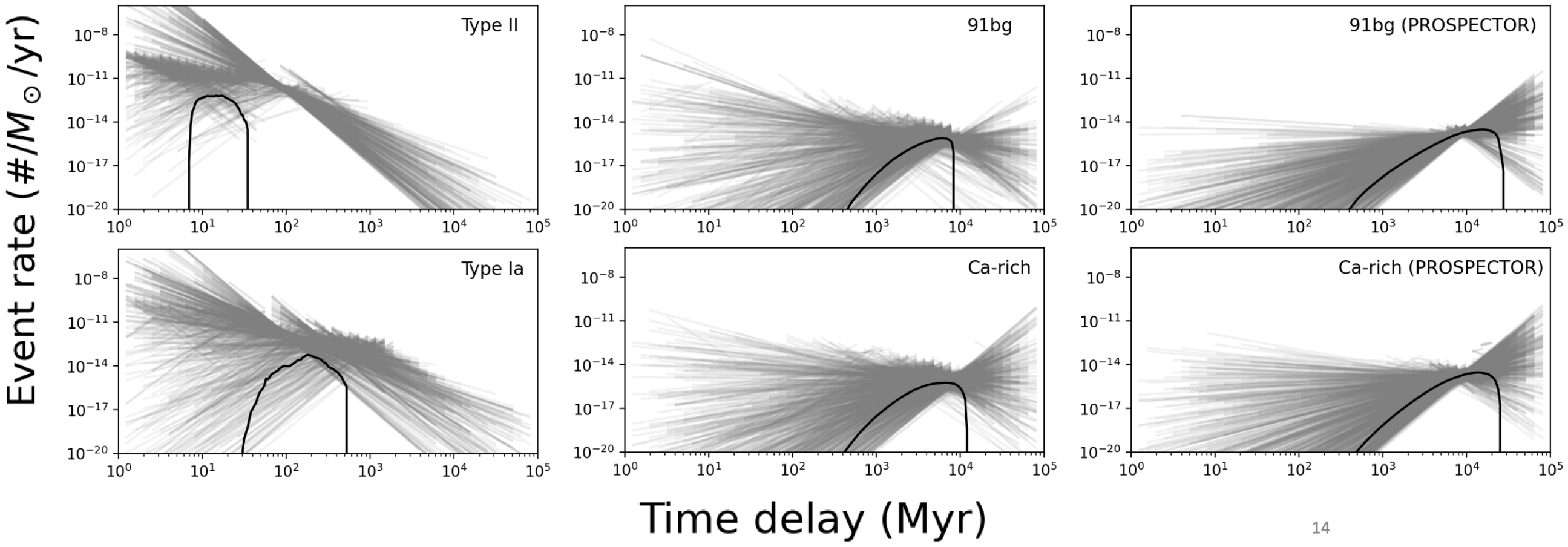

**Figure 6.** Similar to Figure 4, but based on the host galaxies of 4 classes of transients. The left 4 panels have host properties retrieved directly from the CLU catalog, whereas the right 2 panels have host properties from modeling in PROSPECTOR.

Otherwise, we do not see any trend among the subclasses, but this should be investigated further. In particular, De et al. (2020) hypothesizes that Ca-Ia objects are the result of He detonation on higher mass CO WDs, compared to the CO WDs in green or red Ca-Ib/c objects. This prediction should be propagated to predictions of delay times and host properties for these subclasses to compare with observed host properties, but an increase in sample size and spectroscopic followup for identification is needed.

This paper has investigated the global host environments of Ca-rich gap transients, specifically focusing on the mass and SFR of their entire host galaxies. However, the curiously large offsets of these transients from their hosts suggest that the local host properties could also play a significant role (as has been investigated for SNe Ia by Nugent et al. 2024). The host environment is particularly important if the progenitor systems formed in such remote locations. Deep imaging of the local stellar population at the transient site would enable analysis of the immediate environment, providing insight into possible progenitor origins.

The analysis of the hosts of Ca-rich gap transients and 91bg-like SNe demonstrates a plausible link between these two classes of transients. The similarity of host environment and delay times is not a conclusive link between these two classes, but it is a necessary ingredient to motivate further analysis of the similarities. Since 91bg-like SNe are a peculiar subclass of SNe Ia, where the progenitor is an exploding white dwarf in a binary, this lends support to the hypothesis that Ca-rich gap transients also involve a WD in a binary system. In particular, it has been suggested that 91bg-like SNe result from the merger of a CO WD and a He WD (Pakmor et al. 2010; Crocker et al. 2017; Barkhudaryan et al. 2019; Panther et al. 2019). Therefore, this merger channel should be investigated in the context of Ca-rich gap transients through detailed hydrodynamical simulations, as has been done in Morán-Fraile et al. (2024), and comparison to yields and light curves, as in Jacobson-Galán et al. (2021).

## Acknowledgments

We thank Wren Suess for advice on using PROSPECTOR and Joshua Speagle for useful discussion concerning the DYNESTY sampling tool used by PROSPECTOR. This research benefited from interactions enabled by the Gordon and Betty Moore Foundation through grant GBMF5076.

W.J.-G. is supported by NASA through Hubble Fellowship grant HSTHF2-51558.001-A awarded by the Space Telescope Science Institute, which is operated for NASA by the Association of Universities for Research in Astronomy, Inc., under contract NAS5-26555.

Based on observations obtained with the Samuel Oschin 48 inch Telescope at the Palomar Observatory as part of the Zwicky Transient Facility project. ZTF is supported by the National Science Foundation under grant No. AST-1440341 and a collaboration including Caltech, IPAC, the Weizmann Institute for Science, the Oskar Klein Center at Stockholm University, the University of Maryland, the University of Washington, Deutsches Elektronen-Synchrotron and Humboldt University, Los Alamos National Laboratories, the TANGO Consortium of Taiwan, the University of Wisconsin at Milwaukee, and Lawrence Berkeley National Laboratories. Operations are conducted by COO, IPAC, and UW. For publications using products from Phase-II of the survey (taken on or after 2020 December 1), please include this text: Based on observations obtained with the Samuel Oschin Telescope 48 inch and the 60 inch Telescope at the Palomar Observatory as part of the Zwicky Transient Facility project. ZTF is supported by the National Science Foundation under grants No. AST-1440341 and AST-2034437 and a collaboration including current partners Caltech, IPAC, the Weizmann Institute for Science, the Oskar Klein Center at Stockholm University, the University of Maryland, Deutsches Elektronen-Synchrotron and Humboldt University, the TANGO Consortium of Taiwan, the University of Wisconsin at Milwaukee, Trinity College Dublin, Lawrence Livermore National Laboratories, IN2P3, University of Warwick, Ruhr University Bochum, Northwestern University and former partners the University of Washington, Los Alamos National Laboratories, and Lawrence Berkeley National Laboratories. Operations are conducted by COO, IPAC, and UW.

This publication makes use of data products from the Wide-field Infrared Survey Explorer, which is a joint project of the University of California, Los Angeles, and the Jet Propulsion Laboratory/California Institute of Technology, funded by the National Aeronautics and Space Administration.

This publication also makes use of data products from NEOWISE, which is a project of the Jet Propulsion Laboratory/California Institute of Technology, funded by the Planetary Science Division of the National Aeronautics and Space Administration.

This publication makes use of data products from the Two Micron All Sky Survey, which is a joint project of the University of Massachusetts and the Infrared Processing and Analysis Center/California Institute of Technology, funded by the National Aeronautics and Space Administration and the National Science Foundation.

Funding for the Sloan Digital Sky Survey IV has been provided by the Alfred P. Sloan Foundation, the U.S. Department of Energy Office of Science, and the Participating Institutions.

SDSS-IV acknowledges support and resources from the Center for High Performance Computing at the University of Utah. The SDSS website is www.sdss.org.

SDSS-IV is managed by the Astrophysical Research Consortium for the Participating Institutions of the SDSS Collaboration including the Brazilian Participation Group, the Carnegie Institution for Science, Carnegie Mellon University, Center for Astrophysics—Harvard & Smithsonian, the Chilean Participation Group, the French Participation Group, Instituto de Astrofísica de Canarias, The Johns Hopkins University, Kavli Institute for the Physics and Mathematics of the Universe (IPMU)/University of Tokyo, the Korean Participation Group, Lawrence Berkeley National Laboratory, Leibniz Institut für Astrophysik Potsdam (AIP), Max-Planck-Institut für Astronomie (MPIA Heidelberg), Max-Planck-Institut für Astrophysik (MPA Garching), Max-Planck-Institut für Extraterrestrische Physik (MPE), National Astronomical Observatories of China, New Mexico State University, New York University, University of Notre Dame, Observatário Nacional/MCTI, The Ohio State University, Pennsylvania State University, Shanghai Astronomical Observatory, United Kingdom Participation Group, Universidad Nacional Autónoma de México, University of Arizona, University of Colorado Boulder, University of Oxford, University of Portsmouth, University of Utah, University of Virginia, University of Washington, University of Wisconsin, Vanderbilt University, and Yale University.

This research has made use of the NASA/IPAC Extragalactic Database (NED), which is funded by the National Aeronautics and Space Administration and operated by the California Institute of Technology.

*Facilities:* California Institute of Technology 1.2m Samuel Oschin Telescope (also known as 48inch Schmidt) at Palomar Observatory; NASA/IPAC Extragalactic Database.

*Software:* Astropy (Astropy Collaboration et al. 2013, 2018, 2022), astroquery (Ginsburg et al. 2019), DYNESTY (Skilling 2004, 2006; Speagle 2020; Koposov et al. 2024), PROSPECTOR (Johnson et al. 2021), PYROMETER (P. Behroozi 2025, in preparation), Python-fsps (Conroy et al. 2009; Conroy & Gunn 2010a, 2010b), sedpy (Johnson 2019), NumPy (Harris et al. 2020), matplotlib (Hunter 2007).

## ORCID iDs

Peter Scherbak https://orcid.org/0000-0003-4221-9097
Abigail Polin https://orcid.org/0000-0002-1633-6495
Mansi Kasliwal https://orcid.org/0000-0002-5619-4938
Peter Behroozi https://orcid.org/0000-0002-2517-6446
Dave Cook https://orcid.org/0000-0002-6877-7655
W. V. Jacobson-Galán https://orcid.org/0000-0002-3934-2644

## References

Abdurro'uf, Accetta, K., Aerts, C., et al. 2022, ApJS, 259, 35
Astropy Collaboration, Price-Whelan, A. M., Lim, P. L., et al. 2022, ApJ, 935, 167
Astropy Collaboration, Price-Whelan, A. M., Sipőcz, B. M., et al. 2018, AJ, 156, 123
Astropy Collaboration, Robitaille, T. P., Tollerud, E. J., et al. 2013, A&A, 558, A33
Barkhudaryan, L. V., Hakobyan, A. A., Karapetyan, A. G., et al. 2019, MNRAS, 490, 718
Behroozi, P., Wechsler, R. H., Hearin, A. P., & Conroy, C. 2019, MNRAS, 488, 3143
Bellm, E. C., Kulkarni, S. R., Graham, M. J., et al. 2019, PASP, 131, 018002
Bianchi, L., Conti, A., & Shiao, B. 2014, yCat, 2335, I/335
Bildsten, L., Shen, K. J., Weinberg, N. N., & Nelemans, G. 2007, ApJL, 662, L95

Burgaz, U., Maguire, K., Dimitriadis, G., et al. 2025, A&A, 694, A13
Calzetti, D., Armus, L., Bohlin, R. C., et al. 2000, ApJ, 533, 682
Cardelli, J. A., Clayton, G. C., & Mathis, J. S. 1989, ApJ, 345, 245
Chabrier, G. 2003, PASP, 115, 763
Chakraborty, S., Sadler, B., Hoeflich, P., et al. 2024, ApJ, 969, 80
Conroy, C. 2013, ARA&A, 51, 393
Conroy, C., & Gunn, J. E., 2010a FSPS: Flexible Stellar Population Synthesis, ascl:1010.043
Conroy, C., & Gunn, J. E. 2010b, ApJ, 712, 833
Conroy, C., Gunn, J. E., & White, M. 2009, ApJ, 699, 486
Cook, D. O., Kasliwal, M. M., Sistine, A. V., et al. 2019, ApJ, 880, 7
Crocker, R. M., Ruiter, A. J., Seitenzahl, I. R., et al. 2017, NatAs, 1, 0135
Cutri, R. M., Wright, E. L., Conrow, T., et al. 2013, Explanatory Supplement to the AllWISE Data Release Products, Tech. Rep. 1, IPAC/Caltech
Das, K. K., Kasliwal, M. M., Fremling, C., et al. 2023, ApJ, 959, 12
Davis, T. A., Young, L. M., Crocker, A. F., et al. 2014, MNRAS, 444, 3427
De, K., Kasliwal, M. M., Cantwell, T., et al. 2018, ApJ, 866, 72
De, K., Kasliwal, M. M., Tzanidakis, A., et al. 2020, ApJ, 905, 58
Dessart, L., & Hillier, D. J. 2015, MNRAS, 447, 1370
Dimitriadis, G., Burgaz, U., Deckers, M., et al. 2025, A&A, 694, A10
Dong, Y., Milisavljevic, D., Leja, J., et al. 2022, ApJ, 927, 199
Draine, B. T., & Li, A. 2007, ApJ, 657, 810
Ertini, K., Folatelli, G., Martinez, L., et al. 2023, MNRAS, 526, 279
Filippenko, A. V., Richmond, M. W., Branch, D., et al. 1992, AJ, 104, 1543
Foley, R. J. 2015, MNRAS, 452, 2463
Fontana, A., Santini, P., Grazian, A., et al. 2009, A&A, 501, 15
Foreman-Mackey, D., Sick, J., Johnson, B., et al. 2014, python-fsps: Python bindings to FSPS (v0.1.1), Zenodo, doi:10.5281/zenodo.12157
Freundlich, J., & Maoz, D. 2021, MNRAS, 502, 5882
Frohmaier, C., Sullivan, M., Maguire, K., & Nugent, P. 2018, ApJ, 858, 50
Ginsburg, A., Sipőcz, B. M., Brasseur, C. E., et al. 2019, AJ, 157, 98
Goodman, J., & Weare, J. 2010, Communications in Applied Mathematics and Computational Science, 5, 65
Graham, M. J., Kulkarni, S. R., Bellm, E. C., et al. 2019, PASP, 131, 078001
Greggio, L. 2005, A&A, 441, 1055
Haario, H., Saksman, E., & Tamminen, J. 2001, Bernoulli, 7, 223
Hakobyan, A. A., Barkhudaryan, L. V., Karapetyan, A. G., et al. 2020, MNRAS, 499, 1424
Harris, C. R., Millman, K. J., van der Walt, S. J., et al. 2020, Natur, 585, 357
Heringer, E., Pritchet, C., & van Kerkwijk, M. H. 2019, ApJ, 882, 52
Hunter, J. D. 2007, CSE, 9, 90
Irani, I., Prentice, S. J., Schulze, S., et al. 2022, ApJ, 927, 10
Jacobson-Galán, W. V., Margutti, R., Kilpatrick, C. D., et al. 2020a, ApJ, 898, 166
Jacobson-Galán, W. V., Margutti, R., Kilpatrick, C. D., et al. 2021, ApJL, 908, L32
Jacobson-Galán, W. V., Polin, A., Foley, R. J., et al. 2020b, ApJ, 896, 165
Jacobson-Galán, W. V., Venkatraman, P., Margutti, R., et al. 2022, ApJ, 932, 58
Johnson, B. D., 2019 Prospector: Stellar population inference from spectra and SEDs, ascl:1905.026
Johnson, B. D., Leja, J., Conroy, C., & Speagle, J. S. 2021, ApJS, 254, 22
Kasliwal, M. M., Kulkarni, S. R., Gal-Yam, A., et al. 2012, ApJ, 755, 161
Koposov, S., Speagle, J., Barbary, K., et al. 2024, joshspeagle/dynesty: v2.1.4, Zenodo, doi:10.5281/zenodo.12537467
Leibundgut, B., Kirshner, R. P., Phillips, M. M., et al. 1993, AJ, 105, 301
Leja, J., Carnall, A. C., Johnson, B. D., Conroy, C., & Speagle, J. S. 2019, ApJ, 876, 3
Leja, J., Johnson, B. D., Conroy, C., van Dokkum, P. G., & Byler, N. 2017, ApJ, 837, 170
Leroy, A. K., Sandstrom, K. M., Lang, D., et al. 2019, ApJS, 244, 24
Li, W., Leaman, J., Chornock, R., et al. 2011, MNRAS, 412, 1441
Lunnan, R., Kasliwal, M. M., Cao, Y., et al. 2017, ApJ, 836, 60
Lyman, J. D., Levan, A. J., Church, R. P., Davies, M. B., & Tanvir, N. R. 2014, MNRAS, 444, 2157
Mainzer, A., Bauer, J., Grav, T., et al. 2011, ApJ, 731, 53
Maoz, D., Mannucci, F., & Brandt, T. D. 2012, MNRAS, 426, 3282
Martin, D. C., Fanson, J., Schiminovich, D., et al. 2005, ApJL, 619, L1
Masci, F. J., Laher, R. R., Rusholme, B., et al. 2018, PASP, 131, 018003
Milisavljevic, D., Patnaude, D. J., Raymond, J. C., et al. 2017, ApJ, 846, 50
Morán-Fraile, J., Holas, A., Röpke, F. K., Pakmor, R., & Schneider, F. R. N. 2024, A&A, 683, A44
Nugent, A. E., Polin, A. E., & Nugent, P. E. 2024, arXiv:2304.10601
Pacifici, C., Oh, S., Oh, K., Lee, J., & Yi, S. K. 2016, ApJ, 824, 45
Pakmor, R., Kromer, M., Röpke, F. K., et al. 2010, Natur, 463, 61
Panther, F. H., Seitenzahl, I. R., Ruiter, A. J., et al. 2019, PASA, 36, e031
Perets, H. B., & Beniamini, P. 2021, MNRAS, 503, 5997
Perets, H. B., Gal-Yam, A., Mazzali, P. A., et al. 2010, Natur, 465, 322
Perrefort, D., Zhang, Y., Galbany, L., Wood-Vasey, W. M., & González-Gaitán, S. 2020, ApJ, 904, 156
Polin, A., Nugent, P., & Kasen, D. 2019, ApJ, 873, 84
Polin, A., Nugent, P., & Kasen, D. 2021, ApJ, 906, 65
Schlafly, E. F., & Finkbeiner, D. P. 2011, ApJ, 737, 103
Senzel, R., Maguire, K., Burgaz, U., et al. 2025, A&A, 694, A14
Shen, K. J., & Bildsten, L. 2009, ApJ, 699, 1365
Simonian, G. V., & Martini, P. 2017, MNRAS, 464, 3920
Skilling, J. 2004, AIPC, 735, 395
Skilling, J. 2006, BayAn, 1, 833
Skrutskie, M. F., Cutri, R. M., Stiening, R., et al. 2006, AJ, 131, 1163
Speagle, J. S. 2020, MNRAS, 493, 3132
Tauris, T. M., Langer, N., & Podsiadlowski, P. 2015, MNRAS, 451, 2123
Touchard-Paxton, C.-G., Frohmaier, C., Pursiainen, M., et al. 2025, MNRAS, 537, 1015
Turatto, M., Benetti, S., Cappellaro, E., et al. 1996, MNRAS, 283, 1
Waldman, R., Sauer, D., Livne, E., et al. 2011, ApJ, 738, 21
Wright, E. L. 2006, PASP, 118, 1711
Wright, E. L., Eisenhardt, P. R. M., Mainzer, A. K., et al. 2010, AJ, 140, 1868
Zapartas, E., de Mink, S. E., Izzard, R. G., et al. 2017, A&A, 601, A29
Zenati, Y., Perets, H. B., Dessart, L., et al. 2023, ApJ, 944, 22
Zenati, Y., Toonen, S., & Perets, H. B. 2019, MNRAS, 482, 1135
Zheng, Z., & Ramirez-Ruiz, E. 2007, ApJ, 665, 1220